\documentclass[12pt]{article}

\usepackage[T1]{fontenc}
\usepackage[utf8]{inputenc}
\usepackage{amsmath, amssymb, amsfonts, bm}
\usepackage{graphicx,psfrag,epsf}
\usepackage{float}
\usepackage{enumerate}
\usepackage{natbib}

\newcommand{\blind}{0}

\graphicspath{{./}}
\newcommand{\R}{\mathbb{R}}
\newcommand{\N}{\mathbb{N}}

\DeclareMathOperator*{\E}{\mathbb{E}}

\DeclareMathOperator*{\D}{\mathbb{D}}

\begin{document}

\def\spacingset#1{\renewcommand{\baselinestretch}%
		{#1}\small\normalsize} \spacingset{1}
	
%%%%%%%%%%%%%%%%%%%%%%%%%%%%%%%%%%%%%%%%%%%%%%%%%%%%%%%%%%%%%%%%%%%%%%%%%%%%%%
	
	\if0\blind
	{
		\title{\bf Detecting Anomalous Time Series by GAMLSS-Akaike-Weights-Scoring}
		\author{Cole Sodja\thanks{}\hspace{.2cm}\\
			Microsoft, Redmond Wa\\
		}
		\maketitle
	} \fi
	
	\if1\blind
	{
		\bigskip
		\bigskip
		\bigskip
		\begin{center}
			{\LARGE\bf Title}
		\end{center}
		\medskip
	} \fi

\bigskip
\begin{abstract}
An extensible statistical framework for detecting anomalous time series including those with heavy-tailed distributions and non-stationarity in higher-order moments is introduced based on penalized likelihood distributional regression. Specifically, generalized additive models for location, scale, and shape are used to infer sample path representations defined by a parametric distribution with parameters comprised of basis functions. Akaike weights are then applied to each model and time series, yielding a probability measure that can be effectively used to classify and rank anomalous time series. A mathematical exposition is also given to justify the proposed Akaike weight scoring under a suitable model embedding as a way to asymptotically identify anomalous time series. Studies evaluating the methodology on both multiple simulations and real-world datasets also confirm that high accuracy can be obtained detecting many different and complex types of shape anomalies.
Both code implementing GAWS for running on a local machine and the datasets referenced in this paper are available online.
\end{abstract}

\noindent%

\spacingset{1.45}
\section{Introduction}
Univariate time series anomaly detection involves identifying rare or otherwise unsual points in time for a given time series. A related but fundamentally different problem is the detection of anomalous series among a collection as discussed in (Gupta, 2014),
(Hyndman et al, 2015), (Beggel et al, 2018), and recently in (Talagala et al, 2019). 
While a universal definition of an anomalous time series is not agreed upon and depends not only on the data but likely on the domain as well, in this paper anomalous series have the property of being observed extremely infrequently. That is, over a large sample size of multiple series and dense time scale, the probability of observing an anomalous series is in the lower tail.  

The literature discussing the detection of anomalous time series has involved introducing methods to measure deviations of latent shapes or features extracted from time series. That is, given a set of time series, the anomalous series are those having the most relatively unusual shapes in terms of a measure of distance or density. This is a challenging problem as no deterministic labels are readily available indicating which shapes are considered anomalous, thus making the problem unsupervised. Furthermore, many real world time series are nonstationary and can have complex nonlinear shapes and Non-Gaussian distributions. Finally, there is inherent uncertainty in characterizing shapes of time series that should be reflected when attempting to quantify the chance of observing the inferred shapes.  

A time series $y$ can be modeled as a  realization from a random function, that is, given a basis function parameterization  $\theta_{y}$, and associated probability distribution $P_{\theta_{y}}$, view $y \sim P_{\theta_{y}}$ as a random sample,  
called a sample path or just path for short. The vector of basis functions are designed to capture the latent shapes, such as trend, seasonality, amount of autocorrelation,  skewness, and so on, and the distribution reflects the degree of noise in the data generating process. It is assumed that there exists such a distribution $P_{\theta_{y}}$ generating $y$, though it is not known with certainty. 

Taking the statistical association between a time series and the set of models that yield plausible representations, defining anomalous series can then proceed by working with collections of suitably high-dimensional model embeddings capturing all the variation in normal shapes and probability distributions, called the normal or null \textit{model space}, and defining a measure to compare the distance of the best time series alternative model representations to the families in the null model space, as similarly proposed in (Viele, 2001). The anomalous series are then identified as those that have sufficiently small distance. The intuition is that the shape anomalies have an alternative model that is unique or associated to only few other series, thus having a relatively large deviation from the set of plausible models. This is later formalized in Section 3.

The construction of a null and alternative model space determines how paths are characterized, and thus is clearly crucial to the above proposal for defining anomalous series. But how should one proceed with specifying model spaces when the right choice of model families is unknown? This a model selection problem and is a central theme in this paper. Additionally, the choice of measure to compare models is important to ensure desirable statistical properties.    

In this paper a new algorithm is developed to quantify the uncertainty of selected models for representing time series and quantifying deviations from normal models based on relative penalized likelihood distributional regression.  Mixtures of bases are estimated through leveraging generalized additive models for location, scale, and shape (GAMLSS) introduced in (Ribgy, Stasinopoulos, 2005). Comparing models is done based on Kullback-Leibler divergence which is then used to construct an interpretable likelihood score to measure the plausibility of models selected. The likelihood scores are efficiently approximated using Akaike weights, which have a rigorous justification linked to information theory, see (Akaike, 1973) and (Akaike, 1981) or (Burnham, Anderson 2002). Applying Akaike weights hasn't been proposed in the context of anomaly detection to the best of the authors knowledge, though they have been shown to yield competitive results for model averaging in time series forecasting, see (Kolassa 2011). 
The abbreviation {\it GAWS}, for GAMLSS-Akaike-Weights-Scoring is used henceforth to refer to the proposal for integrating these two methodologies for anomaly detection. This proposal is extensible enough to accommodate scoring paths inferred from collections of time series that come from different classes. Furthermore, it can handle series with missing values or irregular time points, nonstationarity, nonlinearity and Non-Gaussian distributions. 

The remainder of this paper proceeds as follows:
Section 2 provides an overview of related research work in detecting anomalous time series. Section 3 gives a statistical formulation of the anomalous time series detection problem through model embeddings, and reviews the model selection problem and connection with information theory. Section 4 outlines the computational details of the GAWS algorithm and reviews GAMLSS penalized likelihood estimation and Akaike weights. Section 5 includes simulation studies describing model performance and sensitivity of GAWS to different data. Section 6 describes choices of basis functions and distributions and shares empirical results benchmarking GAWS on two real world datasets. Finally, Section 7 concludes with a summary and lists future research considerations. 

\section{Related Work}
Statistics research discussing the detection of entire anomalous series has been mentioned in various settings, such as for the identification of failures in space shuttle valves (Salvador and Chan, 2005), server monitoring (Hyndman et al, 2015), finding anomalous start light curves in astronomy (Twomey, et al, 2019), and several other applications. Other research areas, such as functional data clustering and shape detection are also closely related, and some particular papers of note discussing basis function learning and comparisons are summarized below.  

A probabilistic approach for comparing inferred basis functions including in the presence of multiple, unknown heterogeneous subspaces was presented in (Schmutz et al., 2018).  An extension of a Gaussian mixture model for multivariate functional data is developed with functional principal component analysis. While this method is model-based, it is sensitive to the selection of the number of clusters, requires choosing the number of eigenfunctions, and yields basis functions that are not recoverable as no constraints are put on the principal components. 

A general purpose black-box algorithm for finding anomalous time series was proposed in (Beggel et al, 2018). They defined subsequences of time series of a fixed length called “shapelets” to create a feature vector, and then applied a kernel approach to the features using a support vector data description (SVDD) algorithm to detect anomalies of the paths based on lying outside an estimated decision boundary. Experiments showed that this approach was able to discover several types of unusual shapes, particularly for smooth series, but it had problems for time series with high variability. Moreover, understanding the type of shape anomalies that can be discovered via shapelets is challenging. 

An extensible framework for identifying anomalous series in terms of learning and comparing feature vectors was given in (Hyndman et al, 2015). A context-relevant feature space is first constructed, then PCA is applied to reduce dimensionality and finally a multivariate outlier method based on learning an $\alpha$ convex hull is applied to generate a ranking of anomalous series. A nice advantage of this method is that the shape anomalies detected can be controlled through careful construction of interpretable features. However, it can take substantial effort to expand the feature space for new diverse data sets, compared to basis function learning. This approach is used as a benchmark to compare performance to the proposed GAWS methodology. 

A general anomaly detection algorithm termed stray (search and trach anomaly) applicable for high-dimensional data including time series was proposed in (Talagala et al, 2019). This approach also works with extracted features of time series and integrates, though incorporates a density-based measure using extreme value theory together with a modification to k-nearest neighbor searching. This research shows very promising results and is also evaluated as another competing benchmark.

Research that is similar to the GAWS proposal given in this paper working with basis function expansions, using for example penalized B-splines was discussed in (Abraham et al, 2003) for clustering, and in (Tzeng et al, 2018) for measuring dissimilarity working with smoothing splines. Pairwise comparison of functions via their spline basis coefficients using metrics such as L2 or dynamic time warping has indeed be shown to yield good performance in discriminating among clusters. In fact, (Tzeng et al, 2018) construct a modified L2-distance that accounts for the uncertainty in  smoothing spline estimates, yielding a measure of dissimilarity of functions that is integrated with a nearest neighbor algorithm to identify clusters and shape anomalies. Experiments showed that their approach can be preferable compared to model-based clustering and is capable of discovering different types of anomalies. However, the authors only consider learning a conditional expectation using a natural spline, and extending their proposal to ranking anomalies based on conditioning on other moments like the variance or skewness is not mentioned. Thus, the GAWS methodology presented in this paper can be viewed as a generalization of this work, while also offering a theoretically justified alternative measure that yields interpretable likelihood scores. 

Finally, it's also worth mentioning that from the Bayesian perspective, Gaussian processes (GP) are fully probabilistic and yield interpretable bases in terms of choice of reproducing kernels. GP have been successfully applied for analyzing functional data for functional clustering and classification tasks, e.g. see (Shi, Choi, 2011),  as well as detecting anomalous series, see (Pimentel et al, 2013), and (Twomey, et al, 2019). However, estimating GP can be computationally expensive compared to the penalized maximum likelihood approach. Also, while mathematically equivalent, working directly with kernels may not be as natural in some settings as directly specifying bases, and thus from this perspective, GAMLSS offers a frequentist extension to smoothing splines and complimentary alternative to GP learning.

\section{Anomalous Time Series Detection}
A formal statistical framework is proposed in this section for defining anomalous entire time series motivated by model embeddings and an information theoretic measure. A review is also given justifying the use of information criterion measures for model selection.

\subsection{Model Embedding Formulation} 
The perspective is to treat each univariate time series as a sample path coming from an existing but unknown probability distribution that has basis functions as parameters capturing the location, scale and shape of the distribution, or more succinctly shapes. For the remainder of the paper, $y$ will denote a real-valued path over a continuous time interval $T$ coming from a probability distribution $P_{\theta_{y}} $, where $\theta_{y}$ is a vector valued function over $T$ representing the shapes; when the particular association with a given time series is not needed the subscript will be dropped and the notation $\theta$ used to represent shapes. 

It is assumed that the paths can be embedded into a suitably large but finite set of models consisting of distribution families and basis function classes such that the basis functions well approximate the  shapes. Formally, if the paths live in an infinite dimensional space then there exists a countably dense subspace where the shapes have a pointwise convergent representation consisting of finitely many orthogonal basis functions. While this assumption may seem restrictive, it covers a substantial class of time series that have finite mean and covariance living in spaces with well-behaved finite representations, namely the reproducing kernel Hilbert spaces, 
see (Berlinet,Thomas-Agnan, 2004) for mathematical details on representation theorems for stochastic processes.

Let $ b:T \rightarrow \R $ be a basis function generating a finite dimensional subspace $\mathcal{H}_b = span\{ b_{j}|j=1,...,dim_{b}<\infty\} $, 
and $ \mathcal{B} := \{ b:T \rightarrow \R  \} $ be a class of basis functions that approximate the space of shapes for the collection of paths. Define 
$ \mathcal{F} := \{ \mathbf f := (P_{\mathbf \Theta},\mathcal{H}_b)|b \in \mathcal{B}\} $ to be the class of model families, where the elements are  different distribution families and orthogonal basis functions. Let 
$ \mathcal{M}_{\mathbf f } := \{ \mathbf M_{\mathbf f } :=(\mathbf f, \mathbf{a}_{b},|\mathbf f \in \mathcal{F} \} $, where $ \mathbf{a}_{b}:\N \rightarrow \R $ is a vector of coefficients dependent on basis function $b$. Define 
$ \mathcal{M} := \bigcup_{\mathbf f} \mathcal{M}_{\mathbf f} $, referred to as the model space, with elements called models.

The above construction produces a finite model space as an idealization of the space of unknown distributions $P_{\theta_{y}}$. It is further taken that this model space can be decomposed as
$ \mathcal{M} = \mathcal{M}_{0} \cup \mathcal{M}_{a} $, where each $ P_{\theta} \in \mathcal{M}_{0}$ is associated with a subset of paths with designated "normal shapes", and 
$ \mathcal{M}_{a}:= \mathcal{M} \setminus  \mathcal{M}_{0}$ contain models from the "shape anomalies". Given a new path $y$ and uncertainty about its model $P_{\theta_{y}}$, to assess if $y$ is anomalous, that is, if it has shape anomalies, a measure $\pi$ is introduced to quantify how much $P_{y}$ deviates from models in $\mathcal{M}_{0}$. 
The values associated with this measure $\pi_{y}$ are deemed anomaly scores or simply scores. Formally, given a sufficiently small threshold value $\alpha$, a binary classification of $y$ being assigned as an anomalous path is given based on 
$\pi_{y} :=\pi(P_{y},\mathcal{M}_{0} ) < \alpha$. A ranking of which paths are considered most anomalous also readily follows based on ordering scores in ascending order. 
While there are many different choices of measures for comparing distributions, here the Kullback-Leibler divergence is used to construct an interpretable and proper relative likelihood score.

Framing the identification of anomalous paths based on model embeddings and a KL-divergence measure follows a similar approach as given in (Viele, 2001), where the author attempts to quantify how close functional models are to an idealized data generating process for evaluating lack of fit. It is perhaps worth noting that the idea of embedding a sample path representation into a large family of models comprised of both normal and  alternative basis functions and then using KL-divergence to define anomaly scores in this paper was not motivated by the work of Viele but rather inspired from the philosophy of multimodel inference as discussed in (Burnham, Anderson 2002). Indeed, the perspective taken here is that model selection should be central to how inference is carried out, including for the identification of anomalies. Arguably, anomalies can simply be interpreted as samples that arise from an alternative distribution that does not belong to the class of distributions entertained for the given data.

\subsection{Model Selection via Expected KL-Divergence}
Recall that KL-divergence is given by equation (1):
\begin{equation}
D^{KL}(P_{M}, P_{N}) := 
\int P_{M}[y]*log[P_{M}[y]] - 
\int P_{M}[y]*log[P_{N}[y]]
\end{equation}

The choice of KL-divergence as an appropriate measure of closeness between an alternative probability distribution and its data generating process is warranted both for its nice interpretation as well desirable asympototic properties. Specifically per (Berk, 1966), while not a proper metric, it can be shown that it measures the long term loss involved from
using the wrong model given the data, and thus, is very natural to use for model selection. 

In practice, both the true data generating process (again assuming it exists) as well as the null model space $ \mathcal M_0$ comprised of the normal shapes are unknown and thus must be estimated relative to the data. 
Introduce the notation 
$\mathcal Y_{N}$ to be a collection of N sample paths generated from $P_{\theta_{y}}$. In the literature this data is referred to as functional data, and inference of such data called functional data analysis, see (Ramsay, Silverman, 2005) for a classic reference. 
Let $ \hat{\theta}_{0y} $ denote an estimate of the shapes for a specific path $y$ for some model $P_{\hat{\theta}_{0y}} \in \mathcal M_0$, and 
$ \hat{\theta}_{0} $ denote an arbitrary estimate. Even for a single path, there are potentially many models given there is uncertainty, and in Section 4 an approach using GAMLSS will be outlined to generate the $P_{\hat{\theta}_{0}}$.

The initial measure of interest to quantify the average relative loss from selecting models with normal shapes to represent the data generating process conditional on the sample paths is given by 
Equation (2):
\begin{equation}
D^{KL}_{\mathcal Y_{N}} := \inf_{ \hat{\theta}_{0} }
\frac{1} {N} \sum \limits_{y \in \mathcal Y_{N}, \hat{\theta}_{0y} } 
D^{KL}(P_{\theta_{y}}, P_{\hat{\theta}_{0y}})
\end{equation} 

Note that $D^{KL}_{\mathcal Y_{N}}$ is a statistic, dependent on the sample size N and variation among the paths $y \sim P_{\theta_{y}}$. Understanding its distribution is of primary importance. Define  
$\D^{KL}_0 := \E_{ \mathcal Y_{N}}[ D^{KL}_{\mathcal Y_{N}}] $ to be the expected value over the null model space.  Due to a theorem given in (Berk 1966), assuming certain boundedness conditions hold, and
$P_{\theta_{y}} \in \mathcal M_a$ then as N increases the sampling distribution of $D^{KL}_{\mathcal Y_{N}}$ will be a point mass at $\D^{KL}_0$. This is an extremely valuable theoretical result as it ensures that eventually with enough data and a suitable model space the anomalous paths can be detected with arbitrary high probability in terms of the relative average KL-divergence. 

While theoretically justified per the above discussion, it is difficult to directly work with KL-divergence as again the true data generating process cannot be compared against. A model selection procedure that would result in yielding at least an unbiased estimate of $\D^{KL}_0$ would be desirable. This is where the Akaike information criterion (AIC) is useful.

Recall that the penalized negative log likelihood of a model $\mathbf M $ with $\nu_{\mathbf M}$ effective degrees of freedom conditional on a path $y$ is given by Equation (3):
\begin{equation}
NLL_{pen, y}[\mathbf M;\lambda] := -2  \ell[\mathbf M;y]+ \lambda[\nu_{\mathbf M},y] 
\end{equation}
where $\ell $ is the log likelihood function associated with its distribution family $P_{\bm M}$, and 
$ \lambda[\nu_{\mathbf M},y] $ is a penalty; e.g. for AIC, $\lambda =2\nu_{\mathbf M}$, and for BIC,  $\lambda =log[length(y)]\nu_{\mathbf M}$.  

It can be shown that the AIC computed across all paths $y \in \mathcal Y$ is an unbiased estimator of $\D^{KL}_0$, and hence, given N is large enough,  minimizing AIC is equivalent to working with the average relative KL loss $D^{KL}_{\mathcal Y_{N}}$, see (Akaike, 1973) or (Burnham, Anderson 2002).

Choosing the estimate that minimizes AIC won't necessarily result in identifying the true model $P_{\theta_{y}}$, however, utilizing the BIC guarantees this asymptotically, see (Rao, Wu, 1989). However, BIC is not asymptotically efficient, and so for small sample sizes relying on it for model selection can result in underfitting. Therefore, it is important to carefully choose a penalty on the model complexity in accordance with the data and task; this issue is later revisited in the simulations section. 

A particularly nice property of AIC or BIC is that it they can be used to compare different choices of distributions and basis functions, as long as the likelihood is computed on the same time series.

Use of the AIC or BIC for model selection does have its limitations in practical applications. For example, for univariate time series with short sample sizes a modification is often used to correct for bias, but this doesn’t correct for instable parameter estimates. Also, it’s known that AIC tends to favor more complex models than what is needed, whereas relying on BIC for selecting the correct complex model when sample sizes are even modest may fail. While other penalty functions beyond AIC and BIC can be considered, it still is not clear which information criterion is better suited for each particular data set. Probably the biggest limitation is that information criterion-based measures are relative and tell you nothing about the accuracy of the models. Not including enough model families will bias inference, as is later demonstrated through simulations. As always, it is recommended to employ careful prior information when settling on a collection of model families, and perform diagnostic checks where possible to validate model reasonableness.

\section{GAMLSS-Akaike-Weights-Scoring}
The GAMLSS-Akaike-Weights-Scoring (GAWS) algorithm is introduced in this section as a solution for either binary classification or ranking of anomalous series within a collection. The algorithm is designed to operate on different classes of univariate discrete or continuous time series, where there are sufficiently many high frequency series per each class so that basis function learning is feasible. While it is possible to apply the algorithm to a collection of sparse functional series through a modification using mixed GAMLSS and specifying hierarchies, the implementation described in this paper performs parameter estimation per each time series separately. 

Analogous to feature-based models, or selection of kernels for GP learning, a particular choice of both distribution families and basis functions must be made. 
Naturally it is possible to use a default set of very flexible continuous and discrete distributions with parameters for location, scale, and shape consisting of generic bases, such as penalized splines or Fourier expansions for periodic series, but additional customization reflecting the nuances of a given domain is likely required to optimize both computational performance and accuracy.

A nice feature of the GAWS algorithm is that initialization of the null and alternative model spaces, scoring and detecting anomalous series, and updating the model spaces are separate components that can all be parallelized. From this perspective, the GAWS algorithm offers a solution that can scale even on massive sets of time series by leveraging cloud computing.  

Before diving into implementation details of the proposed GAWS algorithm, it is instructive to review GAMLSS penalized likelihood estimation and the justification for applying Akaike weights.

\subsection{GAMLSS Penalized Likelihood Estimation}
Generalized additive models for location, scale, and shape (GAMLSS) introduced in (Ribgy, Stasinopoulos, 2005) is a powerful toolbox for building and comparing semiparametric models for the purpose of distributional regression, complimenting other nonparametric methodologies such as quantile regression and Gaussian processes regression. In short, GAMLSS extends generalized additive models to allow fitting probability distributions both inside as well as outside the exponential family. While various estimation procedures have now been implemented,  penalized likelihood estimation will be the focus. 

Estimation involves finding all parameters of a specified distribution conditional on regressors, including parameters of higher-order moments, like variance, skewness and kurtosis. Since GAMLSS are additive models, they support fitting flexible basis functions capable of learning complex shapes while still yielding interpretable relations. For time series specifically, basis functions provides a natural way to generate a decomposition into meaningful unobservable components, like trend, seasonality, change points, etc. 

GAMLSS maximizes a penalized likelihood function dependent on the chosen distribution. Here the input are univariate time series $ \mathbf{Y}=({y}_{t})|\mathbf \Theta $, conditional on an unknown finite dimensional vector of functions 
$ \mathbf \Theta := ( \theta_{1}, ..., \theta_{p} ) $, with a parametric distribution $ P_{\mathbf \Theta} $. Discrete realizations $Y=(y_{1:N})$ are assumed to be dense taken over a common time grid $T$. Each $y_{i}$ are padded with missing values as needed. Let $ \mathbf X_{T} := (\mathbf{t}_{j})$  be a matrix of common regressors extracted over $T$. For example, this would contain a feature for sequence of time $t$ \( \in\) $T$ , as well as other extracted calendar-based features like hour of day, day of week, and so on. Additional features could easily be incorporated.

For each $ m \in \{ 1, ...,p \} $, a decomposition into finitely many basis functions under chosen link functions $ g_{m}$ are produced, yielding shapes given by Equation (4):
\begin{equation}
\eta_{m,t}  = g_{m}[\theta_{m,t}|\mathbf X_{T}] = \sum_{j=1}^{j=d_{m}} b_{jm}[\mathbf{t}_{j}] 
\end{equation} 

where each basis function can be written in the form $ b_{jm}= B_{\mathbf{t}_{j}m} \mathbf a_{jm} $, $ B_{\mathbf{t}_{j}m} $ is an associated basis matrix dependent on time, and $\mathbf a_{jm} $ is a vector of coefficients subject to a quadratic penalty 
$\mathbf a_{jm}^{T} G_{\mathbf{t}_{j}m} \mathbf a_{jm} $. 

The matrices $G_{\mathbf{t}_{j}m}$ are symmetric with a generalized inverse that is a variance-covariance matrix dependent on a vector of hyperparameters used to define penalties $\mathbf{\lambda}_{jm}$.
The penalty vectors need to be provided, though learning can be done to find an optimal value  maximizing the marginal likelihood or applying generalized cross-validation (GCV), which is the approach taken here.

Then using the RS-algorithm as proposed in (Ribgy, Stasinopoulos, 2005), a penalized log-likelihood is maximized as defined in Equation (5):
\begin{equation}
\ell_{pen}\left(\mathbf \Theta;\mathbf{\lambda}\middle|\mathbf{Y}\right)=\sum_{i=1}^{i=N}\log{\left[P_\mathbf \Theta\left(y_i\right)\right]}-\frac{1}{2}\ \sum_{m=1}^{m=p}\sum_{j=1}^{j=d_{m}}{\lambda_{jm}\mathbf a_{jm}^T G_{\mathbf{t}_{j}m} \mathbf a_{jm}}
\end{equation}

The RS-algorithm effectively takes an iterative approach to solve the penalized likelihood using cycles of iteratively reweighted least squares with Fisher scoring and applying a modified backfitting algorithm (Buja et al, 1989).  

\subsection{Akaike Weights Scoring}
Given there is uncertainty in both the selection and estimation of models, entertaining all plausible models that may have close AICs should be done rather than relying on selecting a single model. Quantifying the relative probability of choosing the minimum AIC would thus be useful in yielding an interpretable score that could be thresholded. This is where Akaike weights are particularly attractive. Equation (6) defines the relative log odds of a model given a path:
\begin{equation}
\Delta_{\mathbf M,y} := NLL_{pen, y}[\mathbf M;\lambda] - min_{\mathbf M} \{  NLL_{pen, y}[\mathbf M;\lambda]  \}
\end{equation}
and Equation (7) gives the {\it Generalized Akaike Weights}:
\begin{equation}
\pi_{ \mathbf M,y} := exp(-.5\Delta_{\mathbf M,y})/\sum \limits_{\mathbf M \in \mathcal M}  exp(-.5\Delta_{\mathbf M,y})
\end{equation} 

Equation (8) then provides a way to compute anomaly scores for a given time series marginalizing across the set of all models defining the normal shapes:
\begin{equation}
\pi_{y} := \sum \limits_{\bm M \in \mathcal M_0}  \pi_{ \mathbf M,y}
\end{equation} 

The scores given by summing the Akaike weights over all possible models representing the normal shapes are interpretable, and computationally straightforward, so as long as the penalized likelihood per model is available.

Following the discussion given in Section 3  results from (Akaike, 1973, 1974, 1981) establishing that the Akaike weights provide a measure of the relative likelihood of a model approximating the lowest expected KL-divergence, and asymptotic results of the posterior distribution of $D^{KL}_{\mathcal Y_{N}}$ per (Berk, 1966), the main result justifying scoring of shape anomalies can now be stated.
\newline

\textbf{Corollary 1: Convergence of Akaike Weight Scores} \newline
Under the assumption that a collection of paths $\mathcal {Y} :=\{y \}$ are generated from a finite set of models $ \mathcal{M} = \mathcal{M}_{0} \cup \mathcal{M}_{a} $ composed of paths defined as normal shapes and shape anomalies respectively, and the boundedness conditions required for the use of the Lebesgue dominated convergence theorem hold per (Berk, 1966), if a sample path $y$ has 
data generating process $P_{\theta_{y}} \in \mathcal{M}_{a}$, then $\pi_y$ will converge to a point mass distribution at 0. Consequently, under the above construction, anomalous series can be detected with high probability given enough samples by ranking or thresholding the Akaike weight scores. 
\newline

\subsection{GAWS Algorithm Design}
The specific implementation of GAWS outlined here is an embarrassingly parallel batch algorithm, intended to leverage distributed computing to construct a model space based on GAMLSS penalized likelihood estimation applied across a set of model families. 

GAWS is designed in a modular way, where creation of the full model space $\mathcal{M}$ can be done offline and incrementally on samples or partitions of time series, yielding per model family basis expansions $(P_{\mathbf \Theta}, \mathcal{H}_b,a_{b}) $ (or optional dimensionality reductions of $a_{b}$) that are consolidated and stored. The proceeding sections outline each step. Refer to Figure 1 for a conceptual illustration. 
\begin{figure}[h]
	\centering
	\includegraphics[width=12cm] {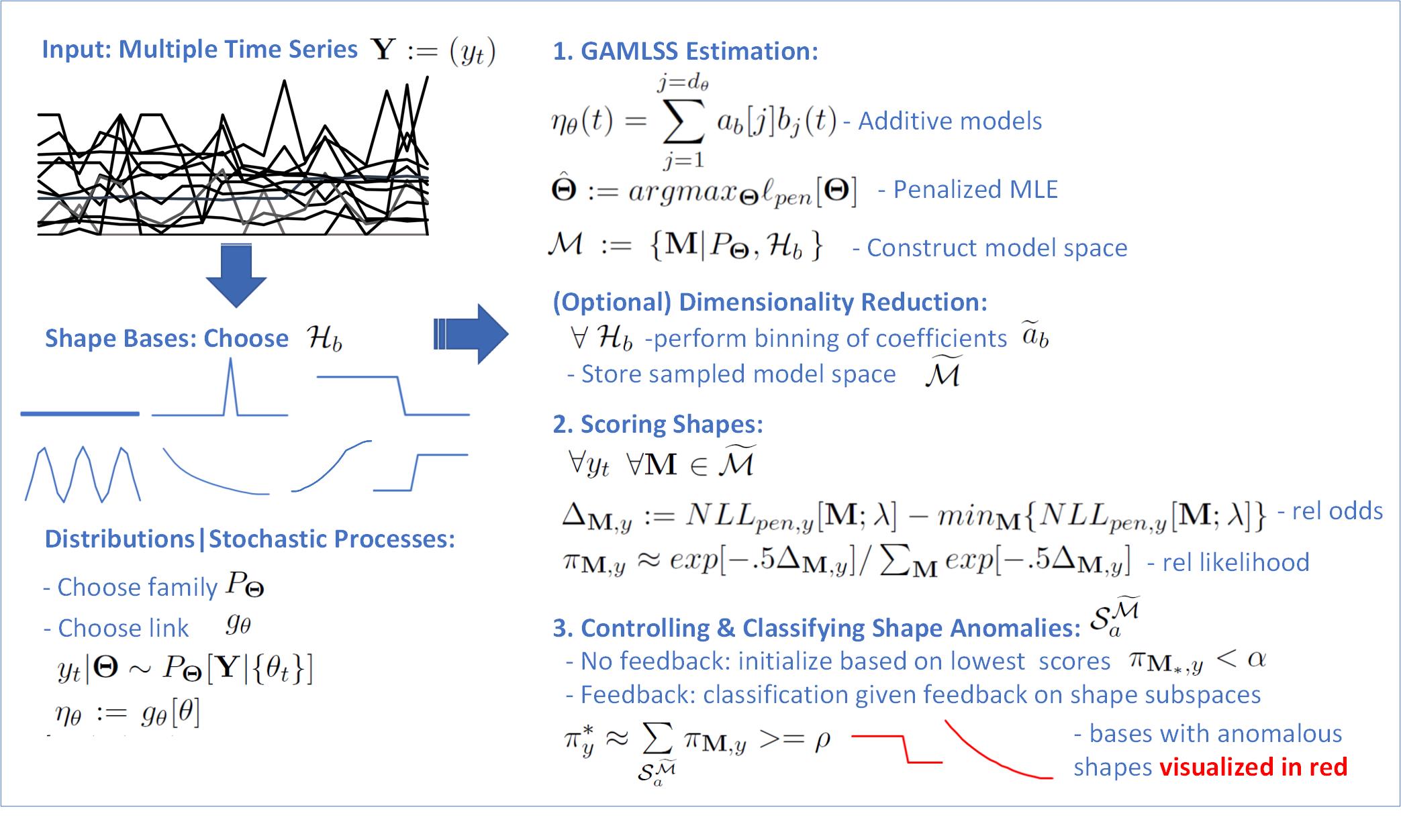}
	\caption{GAWS Framework}
	\label{fig:1}
\end{figure}

\subsection{ Model space Construction}
The GAWS algorithm currently operates by estimating GAMLSS for each univariate time series. To identify a candidate subset of the null model space $\mathcal{M}_0$ associated to a subset of paths with normal shapes, denoted $\mathcal{S}_{0} $, it is important to discriminate among models that are plausible in the sense of having a minimum positive Akaike weight. 

It is hoped that a careful choice of representative orthogonal and constrained basis function classes are chosen up front to reflect the majority of patterns seen. However, to guard against outliers influencing the bases or poorly misspecified model families two control parameters are introduced. 

Let $\alpha$ define the minimum value such that each
Akaike weight $ \pi_{\bm {M}, y} $ must exceed to be considered plausible relative to the union of models $ \{ \bm{M}_y \} $ estimated for $y$ and other models $ \{ \bm{M}_{y_0} \} $. Initialize $\mathcal{M}_0$ to be the set of models $\bm{M}_{y_0}$ satisfying $ \pi_{ \bm{M}_{y_0}, y} >= \alpha $.

While introducing a condition on the minimum Akaike weights to define models $\bm M_0 \in \mathcal{M}_0 $ helps remove some anomalies it is not enough, as given a collection of series arising from multiple complex subspaces it does not differentiate among extremely infrequent model families. Thus, define $N_{min}$ to be the minimum number of unique series  associated with models $\bm M_{0 \bm f} $ needed for a model family $\bm f$. Let $\mathcal F_0$ be the collection of model families satisfying count($\bm {M}_y \in \bm f_0$) $>= N_{min}$, and update $\mathcal{M}_0 = \mathcal{M}_0 \cap \mathcal M_{F_0}$.

A final step removes models associated with anomalous series by thresholding the scores of series $ \pi_y$ relative to 
$\mathcal{M}_0$. This is necessary if there are enough multiple anomalies of the same class that are plausible for one of the model families, which can happen if a model family has basis functions of high complexity. For each series $y_*$ satisfying 
$ \pi_{y_*} < \alpha$, update the set of shape anomalies $\mathcal{S}_{a} = \mathcal{S}_{a} \bigcup \{ y_* \} $,  remove the associated models $ \bm {M}_{y_*}$ from $\mathcal{M}_0$, and initialize the alternative model space $\mathcal{M}_a := \mathcal{M} \setminus \mathcal{M}_0 $.

\subsection{Dimensionality Reduction}
The constructed model space $\mathcal{M}$ might require an excessive amount of storage, particularly when dealing with a massive set of time series or many model families. Moreover, for the purpose of identifying anomalies within each subspace, it's sufficient to cluster or summarize the range of basis coefficients into bins. Thus, a dimensionality reduction step can be employed as necessary to help minimize storage space.

Because of the potential variety of model families, learning a joint distribution across all models is problematic. Therefore, the approach taken here is simple, ignoring the potential gains that could be obtained from applying better multivariate methods by doing it univariately and creating frequency bins for 
$\mathbf{a}_{b}|(P_{\mathbf \Theta},\mathcal{H}_b,\nu_{\mathbf M})$. 
Using histograms to represent basis expansions should also not be problematic  if using orthogonal basis functions.
\newline 
\textbf{Frequency Binning}: \newline 
Initialize $ \widetilde{\mathcal{M}}  = \emptyset $; \newline
Configure bins $\mathbf{I}_b=\left\{[i_{b_k},i_{b_{k+1}})\right\} $;
\newline
$ \forall \mathbf f = (P_{\mathbf \Theta},\mathcal{H}_b,\nu_{\mathbf M}) $ do: \newline
Compute sampling distribution $\widetilde{\mathbf{a}}_{b}:=(\mathbf{I}_b, freq(\mathbf{a}_{b} \in \mathbf{I}_b))$; 
\newline
$\widetilde{\mathcal{M}} := \widetilde{\mathcal{M}} \cup \{ (\mathbf f,\widetilde{\mathbf{a}}_{b}) \}$;

\subsection{Classification and Control of Shape Anomalies}
Given the model space is constructed and a decomposition available, each new series
$y$ can be scored against $\mathcal{M}_0 $ on the fly again employing the lightweight Akaike weights calculations to generate $ \pi_y$. A binary classification can then be performed to designate $y$ anomalous if $\pi_y < \alpha$, or a ranking outputted. 

If feedback becomes available, e.g. domain experts reviewing candidate anomalous series and providing an FP or FN label, a basis function representation is created for the series if feasible, and inserted into the model spaces $\mathcal{M}_0$ and $\mathcal{M}_a$ respectively.

Finally, if identifying particular types of anomalies defined based on a set of a priori shapes considered alarming is the objective, then the alternative scores denoted by $ \pi^*_{y} := 1-\pi_{y}$ can be controlled to achieve a specified expected precision given a parameter $\rho \in (0,1)$. For this problem, it is not necessary to construct the entire model space across all series, but simply compute the score for the null model space of model families against the alternative for each series, thus significantly speeding up computations.

\section{Simulation Study}
A set of synthetic datasets were generated and the performance of GAWS evaluated to assess how well it does finding different types of anomalous series. 

While it would have been nice and indeed faster to assess the GAWS algorithm utilizing cloud computing, only an implementation running for loops on a single machine was executed. A sample size up to 200 time series per simulation run was chosen to yield a reasonable representation while being able to keep run times to hours on a laptop with 4 cores and 16GB RAM running 64 bit Windows. Each time series was measured hourly across 21 days, and composed of location, scale, and shape bases generated from 10 different basis function classes as defined in Subsection 5.2.

The specific distributions, basis functions and parameters chosen in the simulations were based on analyzing real cloud traffic data as outlined in section 6, involving outliers and Non-Gaussian heavy tailed distributions. In particular, the BCCG distribution family was considered to capture changines in the mean and variance, as well as skewness.

The number of time series in each subspace was randomly assigned and constrained to be at least 10; this minimum sample size seemed reasonable to reflect the sampling variation in the parameters considered and frequency of anomalies tested.  

\subsection{Box-Cox-Cole-Green Distribution}
The BCCG distribution as defined in (Cole, Green, 1992) is useful for modeling positive continuous data with excess skewness. It is parameterized via location, scale and shape stochastic processes. \newline 

\textbf{BCCG Density Function} 

BCCG[$y_t | \mu_t, \sigma_t, \nu_t$] $: =
\frac{1} {\sqrt{2\pi} \sigma }
\frac{ y_t^{\nu_t -1} } {\mu_t^{\nu_t}}
exp[- \frac{z_t^2} {2}] $ \newline

where if $\nu_t \ne 0$ then 
$z_t:= [( \frac {y_t} {\mu_t} )^{\nu_t} -1]/(\nu_t \sigma_t)$, otherwise,
$z_t:= log[\frac {y_t} {\mu_t}]/ \sigma_t;$ \newline

\subsection{Simulated Basis Functions and Parameters}
Sample paths are defined based on a composition of additive models under log link functions for the BCCG distribution. Formally, each time series $y_{t}$ are independent, identically distributed samples from $P_{\mathbf \Theta}$, where the function parameters 
$\mathbf \Theta = \{ \theta_t = g^{-1}_\theta[{\eta_\theta}_t] \}$, are given by log canonical link functions $g^{-1}_\theta$, and shape functions $ {\eta_\theta}_t = \sum_{m=1}^{m=M} b_{m,t} $ represented in terms of finitely many prespecified basis functions $b_{m,t}$.

For location parameters, the default basis is constructed from the local level with daily and weekly seasonality simulated from a double seasonal innovations state space model as posed in (Gould et al, 2008). Additional bases include random pulses, autoregressive processes, and linear step functions to capture level shifts.

For scale parameters, constants sampled from a grid of values between .05 to .25 forms a default basis. The random walk with slow growth is also considered to capture shape anomalies in increasing variance. For the BCCG distribution shape parameter, i.e. skewness, a range of constant values randomly sampled between -.5 and .2 is considered, which maintains right-tail skewed data.  

Let $ \epsilon_t \sim N(0, \sigma^2)$ be iid Gaussian noise, where for each time series, $y_t$, \newline $\sigma^2 = cv *(1+ y_1$), where  $cv$ is the coefficient of variation, generated from a truncated log-normal distribution with location log(.05), and scale .25. These priors were chosen to evaluate a range of very low to medium relative variance in the location parameters. The following models form the collection of bases:
\newline

\textbf{Local Level Model} 

$ L_t := L_{t-1} + \alpha \epsilon_t$;

$\alpha$ randomly between 0 to .15.

$L_0$ is sampled from a truncated log-normal with mean log(500), scale .1, 

and range between 350 and 650. \newline

\textbf{Double Seasonal Model} 

Given time periods $m_1:=7$ and  $m_2:=24$, 
$S_t := S_{t-m1}^{(1)} + S_{t-m2}^{(2)}$ ; 

$S_{t}^{(1)} := S_{t-m1}^{(1)} + \gamma_1 \epsilon_t  $ ; 

$S_{t}^{(2)} := S_{t-m2}^{(2)} + \gamma_2 \epsilon_t  $ ; 

where $\gamma_1 , \gamma_2 $ randomly between .001 and .1, 

$S_{0}^{(1)} := L_0*[.27,.25,.24,.21,.12,-.52,-.57]) $ ; 

$S_{0,n}^{(2)} := L_0*\sum_k c_k sin(2 \pi k n/m2) + d_k cos(2 \pi k n/m2)$ ,

where $n = 0, ...,23$, $c_1 = .1, c_2 = -.2, d_1 = -.5, d_2 = -.2$.
\newline

\textbf{Random Pulse} 

$\exists \chi_t \sim $ Bernoulli[$\pi_t$], and
$ \delta_t$, with $ x_t := \delta_t$ if $\chi_t=1$ and 0 otherwise; 

$ \pi_t $ randomly selected between 0 to .01, and 
$\delta_t = L_t r $, 

$r$ randomly between 3 and 6. 
\newline

\textbf{AR(P) Positively Correlated Processes}

$ x_t := \sum_{i=1}^{i=P}{\phi_i x_{t-i}}+\varepsilon_t $; 

$P$ sampled from a zero-adjusted Poisson distribution with location .2

and proportion of zeros of .75, and  $\phi_i $ randomly between .05 and .25. \newline

\textbf{Random Walk with Positive Drift}

$ x_t := x_{t-1} + b + \epsilon_t $; 

where $b = r*L_0$, $r$ randomly chosen between .0001 and .002. \newline

\textbf{Linear Step Functions}

$\exists \tau_1 \le \ ... \le \tau_N,$ and $ \exists \delta_1,..., \delta_N$,

$ x_t := \sum {\delta_i 1_{[ \tau_i, \tau_{i+1})}(t)}+\epsilon_t $;

$N$ randomly set to 0 with probability .9 and 1 otherwise;

$\tau_i$ is a randomly sampled integer between 50 and 450;

Each $\delta_i =  L_i r$ is a random pulse, with $r$ between .3 to .7 or 1.4 to 2. \newline

\subsection{Location, Scale, and Shape Estimates}
Given a time series $y_t$, models are estimated using GAMLSS penalized likelihood as previously detailed, with location parameters consisting of penalized linear B-splines per (Eilers, Marx, 1996) for modeling trend, penalized cubic cyclic splines applied to hour of day and day of week regressors, AR(p), and regressors for pulses. Scale and shape parameters include intercepts without any regressors, and B-splines applied to time. That is, the path shapes 
$\bm \eta_{y} : = (m_y, s_y, \nu_y)$ are given as follows: \newline

$\boldsymbol{Location}$:

$ m_y(t) = \beta_y^{0} + f_{trend}(t) + f_{c1}(x_{t,dow}) + f_{c2}(x_{t,hour})+
\sum_{k=1} ^{k=N} \delta_k\bm 1_{t_k}(t) +\zeta_t$;  

where $ \zeta_t \sim $ ARMA(p,q), $x_{t,dow} \in \{1,2,...,7 \}$, $x_{t,hour} \in \{ 0,1,...,23 \}$, and 

pulses $\delta_k$ occurring at unknown $t_k$, for $k=1, ...,N$.  \newline

$\boldsymbol{Scale}$ and $\boldsymbol{Shape}$:

$ s_y(t), \nu_y(t) = \beta_y^{0} + f_{trend}(t)$, 
where $f_{trend} = \sum{c_i B_{i,n}\left(t\right) ,\ B_{i,1}\left(t\right)=1\ }$ if $t\in[t_i,t_{i+1}]$, and 0 otherwise, and 

$ B_{i,k+1}\left(t\right) = {\frac{t-t_i}{t_{i+k}-t_i}B}_{i,k}\left(t\right)+\ {\frac{t_{i+k+1}-t}{t_{i+k+1}-t_{i+1}}B}_{i+1,k}\left(t\right)\ $ given knots $t_0,t_1,…,\ t_K $. \newline

In addition to the BCCG distribution, Gamma was also tested to assess if the Akaike Weights were lower under misspecification of the distribution.

\subsection{Generating Shape Anomalies}
Real anomalous time series are rare and can take on shapes that have most of the characteristics of the normal time series but with subtle differences or unusual variations, such as periodicity being masked by an extreme temporary shift. The approach taken to simulate shape anomalies was to  randomly sample from multiple time series with random walk or downward shift for location, linear increasing scale, or shape parameter between -1 and -.5, and combine their values to form new composite time series. This yielded shapes that do not probabilistically belong to any of the defined subspaces, and exaggerates the characteristics taken from the normal but less frequent paths. A sample size of anomalies tested was 1, 5, and 10 per simulation. 
See Figure 2 for a plot of the time series having shape anomalies.

\begin{figure}[h]
	\centering
	\includegraphics[width=10cm] {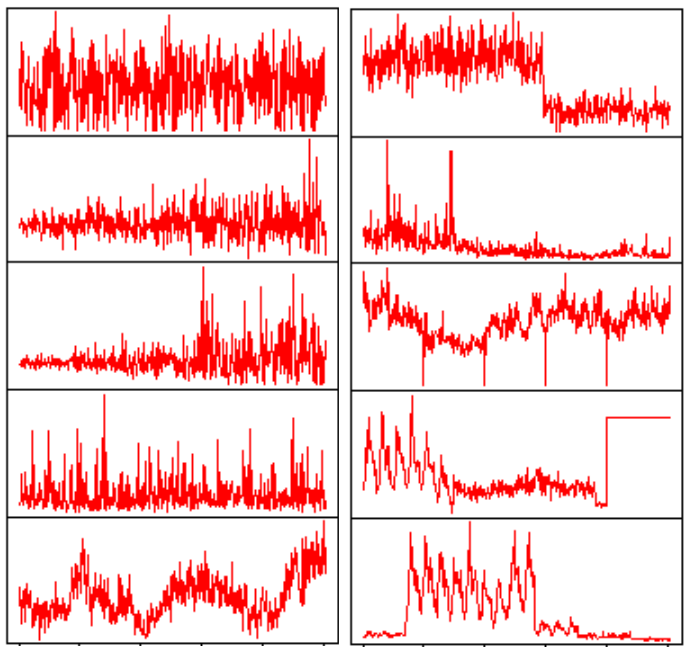}
	\caption{Simulated Shape Anomalies}
	\label{fig:2}
\end{figure}

\subsection{Performance Measure}
The primary measure used to evaluate performance of detecting anomalies is the F score, a weighted average of precision and recall defined by \newline
$ 2*(Precision*Recall)/(Precision+Recall)$, where \newline 
Precision $ =  TP/(TP+FP) $, and 
Recall $ =  TP/(TP+FN) $.
Both precision and recall values are computed over the top 10 ranked unusual time series.

Additionally, a measure of excess rank is introduced: \newline 
if $p$ is the real frequency of anomalies, and $N$ is the number of time series, let $R_N$ be the  position of each anomaly out of $N$ found by sorting the estimated likelihood scores in decreasing order. Then the excess rank is the average difference between frequency and max ranked proportion, i.e.
$E[ p - max(R_N)/N]$. If likelihood scores are properly calibrated then the excess rank should be zero. Note that a weighted average excess rank could also be computed if a measure of rarity of each anomaly was available, but for this simulation each are viewed as equally likely.

\subsection{Benchmarking Performance}
To compare the performance of GAWS in detecting shape anomalies, two other scalable, feature construction based methodologies were considered. Namely time series feature extraction with PCA and multivariate highest density region based anomaly detection proposed in (Hyndman et al, 2015) using the R package anomalous, and the stray algorithm that applies extreme value theory and k-nearest neighbors to classify and rank anomalies working directly with the extracted time series features.

These method are abbreviated {\it hdr} and 
{\it stray} respectively for reporting results.
Note that only the default time series extractor provided by the packages was utilized. It is likely that if custom features were built for the specific simulated data then results would change. 

\subsection{Results}
Normal sample paths were simulated from BCCG processes with the following constraints:
\newline -constant scale/shape split into low values, between .05 and .1 for scale and between -.3 to .2 for shape, and mid values with scale between .1 and .2, and shape between -.5 and -.25, and
\newline -location parameters were generated from a mix of basis function classes, including constant level, multiple seasonality, stationary local level with random pulses, AR(p), and linear upward shifts.

Six different experiments were ran, each generating multiple replications of 200 series during each run, randomly sampling from both the different bases and anomalous series, per each frequency of anomalies tested. 

Experiment 1 consisted of only shapes from a single model family generated from multiple seasonality bases with low and high constant values for scale and shape. It was designed to establish a baseline for relatively simple series.

Experiment 2 consisted of 4 distinct model families from the multiple seasonality, local level random pulses, AR(p), and linear upward shift bases, with 6 different subspaces. This produces a fairly complex mix of time series with different shapes.

Experiment 3 was setup to assess how the methods would perform in the presence of a much larger frequency of anomalies. In addition to the 10 defined anomalies, 10 other anomalies were sampled from a basis consisting of 3 change points and a temporary upward shift. A constant location, with low scale and shape parameters from a BCCG distribution formed the normal series.  

Experiment 4 was designed to test that GAWS with a BCCG distribution family but generic basis function class consisting only of penalized B-splines for all parameters would not overfit. 
Again a constant location, scale and shape model family generated the normal series. 

Experiment 5 is similar to 4, but includes all model families from experiment 2 in the generation of the normal series, as well as an additional local level. 

Finally, experiment 6 was added to show that GAWS could yield very poor results if the wrong model families are entertained. Random walks bases from Gaussian and Student t distribution families  were considered, with again a constant location, scale and shape model family forming the normal time series.

A total of 15 distinct model families were considered running the GAWS algorithm for the main three experiments E1,E2,E3, fitting models for each series including combinations of local level, double seasonality, random pulse, ar(p), random walks, linear shifts, and penalized B-spline trends, using the Gamma and BCCG distribution families. Including additional spurious models was purposely done to evaluate the robustness of Akaike weight scoring. 

Note that time series were re-scaled to have the same mean. This was done because the focus is not on quantifying magnitude differences but rather understanding which relative shapes are unusual.  

Table 1 summarizes the overall average performance for each of the experiments ran. 
Note that the relative F-score is defined as F-score + correction, where 
correction:= $ 
1- 2*\frac{count(anomaly)/10)} {1+count(anomaly)/10} $. The correction is added to account for the maximum possible F-score that can be attained only selecting the top 10. Thus, if there is only 1 anomaly but it is ranked in the top 10 then the relative f-score is 1, instead of an f-score of .182.

\begin{table}[h!]
	\centering
	\includegraphics[width=14cm] {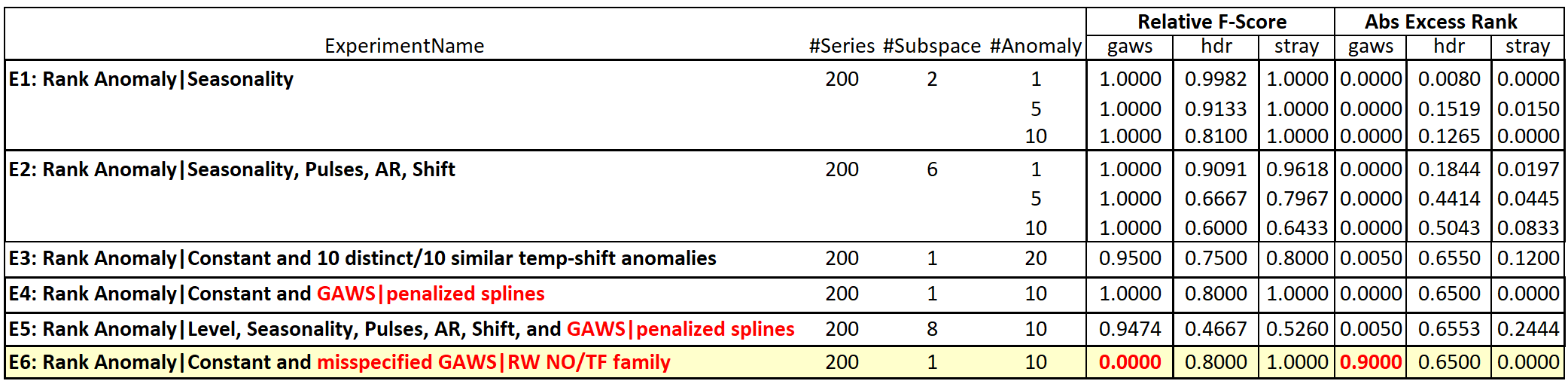}
	\caption{Average Performance per Experiment and Anomaly Frequency}
\end{table}

In summary, across the 5 main experiments E1-E5, GAWS performance often exceeds and is never worse than both the stray and hdr algorithms, and is much better in E2 for higher frequency of anomalies, as well as significantly more accurate for E5. 
GAWS yielded an overall mean relative F-score of .987, a .13 improvement compared to stray and .22 over hdr. In terms of the 
mean absolute excess rank metric, GAWS had a mean of .001, versus stray at .0585, and hdr at .375. In particular, GAWS had perfect ranking except in experiment 5, where it only took 1 additional rank to catch all anomalies.  

Perhaps it is not surprising that GAWS yields such nearly perfect results given the model space constructions are representative of the data, with each model family having adequate samples for GAMLSS to learn basis functions.

As a side note worth calling out, stray is always better than the hdr method, with superior results for the excess rank metric. It does appear that performance for both the hdr and stray algorithms do degrade as the number of anomalies increases in the data. Further experiments would need to be ran to better understand if this is only specific to the type of anomalies and default extracted features considered here.

On the other extreme, GAWS yields inferior results in experiment 6, though this is purely contrived. The GAWS algorithm by default uses the more flexible penalized spline bases, which was shown to yield competitive performance on this simulated data. However, the takeaway from E6 should be that using poor choices of bases can render GAWS useless, which of course is not only the case for this algorithm but any that rely on sensible model or feature choices.   

\section{Real World Applications}
\subsection{Detecting Anomalous Series of Cloud Traffic}
Monitoring traffic for cloud products across multiple data centers and software applications is an important task. As engineers develop and ship new features, update virtual machine configurations, and attempt to improve the cloud experience for customers, inadvertently bugs and other issues can arise and impact performance. Furthermore, because of maintenance, or capacity shortages, traffic can get redirected from the original intended data center to another where excess capacity is available, thus adding noise to the data.

There are too many complex time series for subject matter experts to manually visualize and identify what to investigate, so an automated solution that ranks the most unusual series is useful. Thus, an analysis was initiated to understand how well the GAWS algorithm could perform identifying previously known anomalous series.

A data set consisting of 138 per 30-minute time series with 960 distinct time points reporting the total number of sessions across multiple data centers and different software applications was considered. There were 4 known odd series flagged, and engineers were interested in evaluating which if these could be captured by an anomaly detector looking at the top 10 ranking. An assessment was carried out to comparing rankings from GAWS and stray.

To use the GAWS algorithm, a collection of basis functions and distribution families need to be specified. Performing some exploratory data analysis, and looking at plots for a sample of time series revealed clear multiple and changing seasonality, changing level/slow trend, infrequent spikes, and variability much higher during peak times, see Figure 3. Note that due to Microsoft confidentiality, all references to product information and times are removed in the plots, and actual time series values are transformed while preserving non-negativity and perturbed by adding some Gaussian noise.

\begin{figure}[h]
	\centering
	\includegraphics[width=12cm] {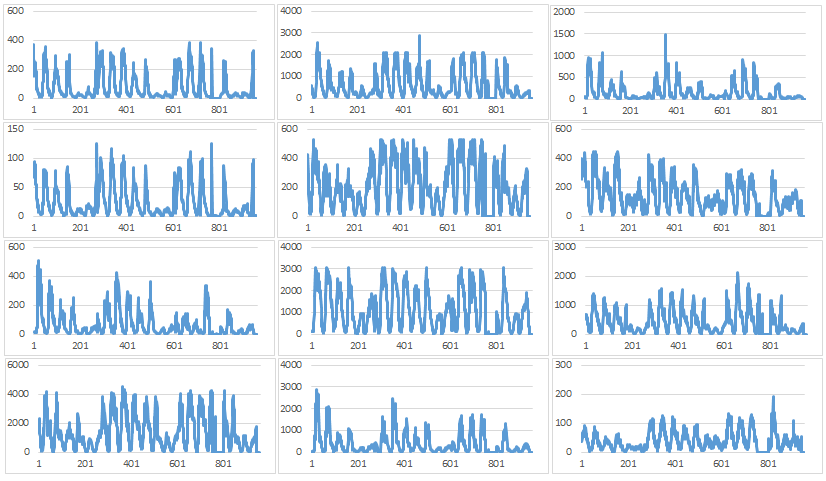}
	\caption{Sample Cloud Traffic Time Series 30-Minute  Intervals}
	\label{fig:2}
\end{figure}

Given the time series were positive skewed large volume counts, several continuous distribution families were entertained including log-normal, log-t, log-skewed-t, gamma, and BCCG with log link. Fitting initial GAMLSS models to the sample of time series with these distribution families was carried out using the following basis functions: for the location parameter, fourier expansions were used to represent seasonality, penalized b-spline of degree 1 captured any changing local level/trend, and binary indicators included to handle significant random pulses.
For the scale parameter,  penalized splines dependent on seasonality and level basis were implemented. For the shape parameter, a constant basis as well as spline conditioned on the level of the series were considered.

Checking a random sample of 20 series confirmed that the model specification with log-t family was reasonable per Figure 4 depicting random fluctuations around center line with few points boarding or outside the 95 percent confidence band of the worm plot, that is, detrended Q-Q plot as described in (Buuren and Fredriks, 2001) applied to the quantile residuals, see (Dunn, 1996). 

\begin{figure}[h]
	\centering
	\includegraphics[width=12cm] {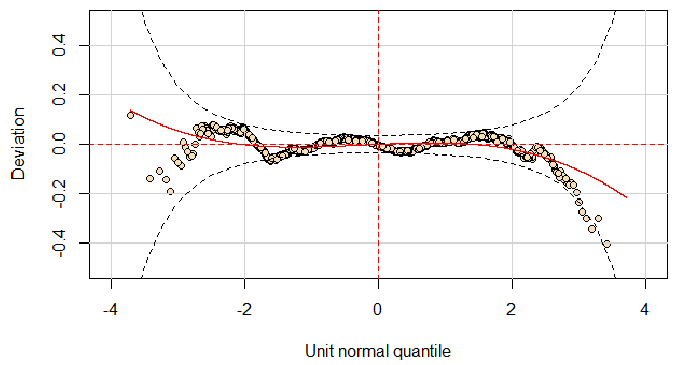}
	\caption{Log-t Distribution Family Worm Plot Quantile Residuals}
\end{figure}

It is also interesting to visualize how the scale changes as a function of the relative peak season, and shape given level. As can be seen in Figure 5, the scale is nonlinearly increasing as a function of season with a maximum occurring between the range of .5 and 1, whereas the shape jumps upward and climbs as the level increases. Being able to model conditional moments in terms of basis functions is easily done leveraging GAMLSS, and is integral to how the GAWS algorithm can detect many types of unusual shapes.

\begin{figure}[h]
	\centering
	\includegraphics[width=12cm] {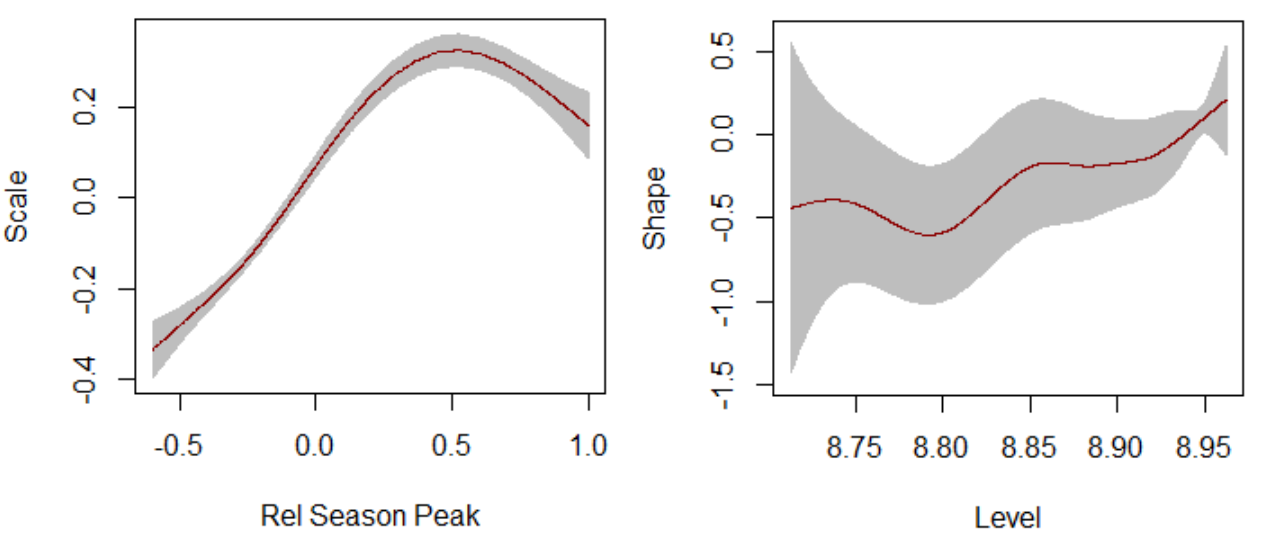}
	\caption{Sample Cloud Traffic Scale and Shape Basis Functions per Time}
\end{figure}

Next, GAWS with the basis functions as defined above and stray were ran, and their rankings outputted and compared, see Table 2. For this data and choice of basis functions, GAWS was able to detect all 4 anomalies in the top 4 ranking as indicated by bold red, with 1 FP for series identifier 130.
The stray algorithm was only able to catch 2 anomalies in the top 5 ranking, and required going to rank 18 to catch all 4. 

\begin{table}[h!]
	\centering
	\includegraphics[width=8cm] {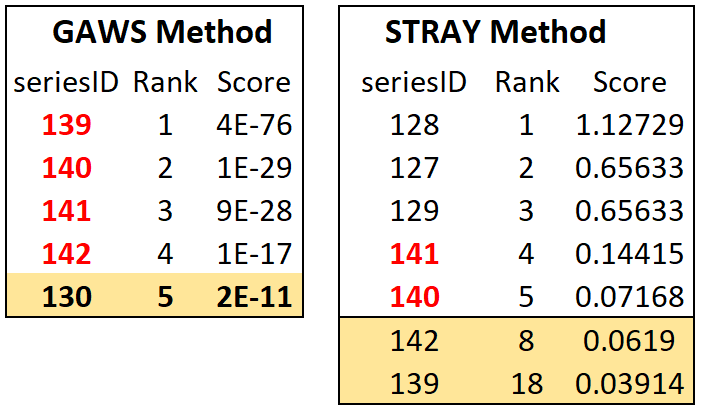}
	\caption{Comparison of Ranking of Cloud Traffic Time Series, with Anomalies in red}
\end{table} 

It is worth noting that stray detected series identifiers 127, 128, 129 as the most anomalous, which were from a specific data center with slightly different seasonal variation that had less noise but otherwise were not remarkable.

Perhaps with custom feature engineering the stray algorithm could have caught all 4 anomalies, though the key takeaway is that GAWS achieved perfect results using a fairly generic class of basis functions.

\subsection{Classifying Anomalous Pedestrian Count Series}
The objective of this study was to show that the GAWS algorithm could successfully identify  specific anomalous series that was detected by the stray method. Specifically, this entailed analyzing a dataset generated from an automated pedestrian count system from sensors in Melbourne, Australia. Tracking unusual counts at different locations and times was used as an indicator of economic health and thus considered an important application.

The dataset consisted of hourly count series taken across the month of Dec 2018 for 43 sensor locations, and is available as a dataframe called ped\_data in the R package stray. 

As discussed in (Talagala et al, 2019), applying the stray method with seven different time series features to this data resulted in finding a particular anomaly on 31-Dec-2018 associated with fireworks for New Year's Eve at the Southbank sensor, see Figure 6.

\begin{figure}[h]
	\centering
	\includegraphics[width=12cm] {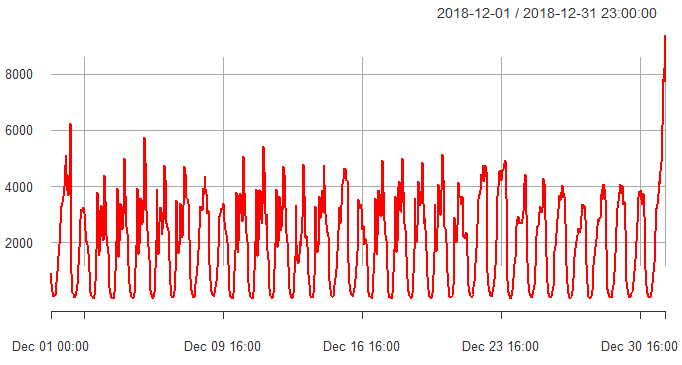}
	\caption{Hourly Pedestrian Counts from Southbank Sensor}
\end{figure}

Looking at the collection of 43 series, 4 series had significantly many missing values for entire portions of time and thus were filtered from consideration for this study. Several remaining series still had some missing values, thus a basis function representation for the location consisting of a penalized spline as a function of time, cyclic splines applied to hour and day of week indicators, and indicator variables to capture spikes were estimated and used to impute missing values. An additional normalization step was performed to ensure each series had the same mean.

Given the normalized and complete series the next step was to create a normal and alternative set of models to discriminate the series from the Southbank shapes. Taking a closer look at the Southbank time series revealed daily and weekly multiplicative seasonality and clear level shift approximately at time index 740, with a ratio between 1.61 and 2.61 comparing to the average from previous days at the same hours. Variance significantly depended on hour of day and weakly on the day of week, though this pattern persisted for many other series. Further, no clear changes were evident in skewness or kurtosis. 

Thus, an alternative model family was introduced to capture the key characteristics of the count series for Southbank and was defined by a log-TF distribution family with location basis functions comprised of multiple seasonality using day of week, hour of day and their interactions, and a temporary upward step function of length at least 3, and constant scale/ shape. A single baseline model family utilized for the other series was a log-TF distribution family with multiple seasonality also using hour and day of week indicator variables for location, and constant scale/shape. 

Since the goal was to assess the relative likelihood of selecting the normal model family versus  alternative with level shift basis functions, running GAWS once per series independently was sufficient, by-passing the construction of a model space and loop comparing each series against all models. Doing this resulted in Southbank series having the lowest relative likelihood of being normal, 
about $ < 7.7*10^{-12} $. Hence, the GAWS algorithm was easily able to detect the largest anomaly consistent with stray.

It is interesting to look at other time series with low Akaike weight scores to assess shapes.
Figure 7 plots highlights in red time series with scores <.0015, and all others in grey. Although the stray method did not have any of these series in the top 6 ranking beyond Southbank, arguably these are very similar in that they also have a level shift for hours on 31-Dec-2018, albeit not as extreme as observed in Southbank. From this viewpoint, GAWS is doing a good job discriminating among other series that do not have this type of pattern. Including more models to capture further differences in shapes could be entertained and 
would be expected to result in even better discrimination among the shapes.

\begin{figure}[h!]
	\centering
	\includegraphics[width=14cm] {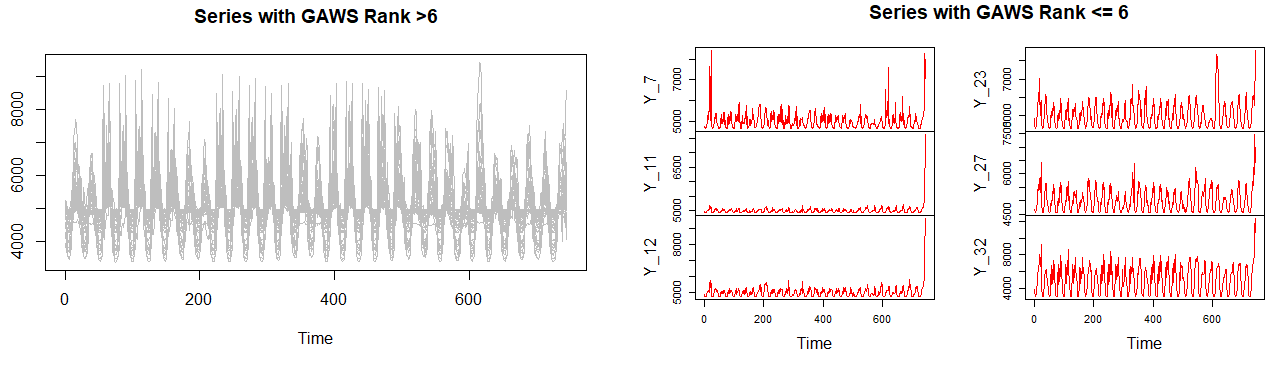}
	\caption{Comparing Pedestrian Count Series Ranked by GAWS Algorithm}
\end{figure}

\newpage

\section{Discussion}
In this paper, the problem of detecting relatively anomalous time series among collections was framed and probabilistically quantified. Viewing a time series as a sample path from an unknown random function, a relative likelihood score is applied to its shapes to establish a measure of rarity. Namely, a shape is identified as anomalous if it is implausible to be represented by models belonging to a sufficiently large class comprised of well-specified probability distribution families and basis functions that capture the typical data generating process. From this perspective, anomalous series have shapes that belong to an alternative, low-dimensional and well-separated subspace.  

A fairly general statistical framework was then formalized for detecting shape anomalies based on model embeddings and Akaike weights following a similar philosophy as discussed in (Viele, 2001). A mathematical justification was also presented, explaining why the proposed Akaike weight scores are at least asymptotically unbiased referencing previous known results connecting KL divergence type measures and information criterion. 

The GAWS algorithm was then introduced as a data-driven solution for ranking and classifying anomalous series, utilizing generalized additive models for location, scale, and shape to produce flexible model embeddings, and computing Akaike weight scores via penalized likelihood. GAWS provides interpretable inference and can be implemented in a scalable way. 

In addition to the supporting asymptotic theory guaranteeing that the Akaike weight scores will converge to a point mass distribution centered at zero for time series with shape anomalies under suitable regularity conditions and proper choice of model families, it was also empirically demonstrated that the proposed GAWS method can yield excellent accuracy in detecting anomalies among complex classes of series. Specifically, looking at both multiple simulations and two real datasets, GAWS achieved very high precision and recall, and always yielded as good and often much better results than other methods, including the recently proposed stray algorithm in (Talagala et al, 2019). However, it is acknowledged that no general claims can be made regarding the performance of GAWS across all possible situations, and certainly further studies are warranted to continue assessing the strengths and limitations of GAWS give particular time series. 

While GAWS provides a powerful toolbox for detection of anomalous time series, it does have some known limitations. The penalized likelihood formulation, while computationally appealing, is restricted to models where the effective degrees of freedom is known. Moreover, there needs to be ample representative data to obtain trustworthy likelihood estimates. It is expected that GAWS would not perform as well dealing with collections of only short time series, where majority are both highly sparse and heterogenous.

Like other feature-based methods, GAWS is only useful if a reasonable set of model families are entertained. Utilizing generic bases such as penalized splines to construct an embedding can work quite well given enough data as was demonstrated in the simulation study, but specifying particular location, scale and shape parameters is likely needed to help reduce both false positives and false negatives in practice. 

Although GAWS was designed for the purpose of anomaly detection, an extension to functional clustering seems plausible, where the location, scale and shape basis expansions together with an information criterion could be utilized to form fuzzy clusters and hence provide an alternative way to generate mixture models via the Akaike weights. Further work is required to assess the feasibility of this capability.

Additional research is also needed to extend the GAWS framework before it can be useful in certain settings. For example, enhancements would need to be made for GAWS to be suitable for streaming data, where both the model space and parameters are dynamically updated. Another important consideration is automatic pooling to learn parameters across groups of series given sparse data, and improve estimation and inference working with hierarchical time series. Integrating and evaluating an alternative estimation procedure, such as boosting for datasets requiring specification of many basis function classes would be worthwhile for both computational speed and practical construction. Incorporating mixtures of model families for complex datasets where unimodal parametric distributions are insufficient would make the GAWS algorithm even more extensible. 
On a final note, it would be interesting to re-frame GAWS within a Bayesian framework, putting a prior on the choice of model space and computing scores via full Bayesian model averaging. This would open up the use of other information criterion measures to explore as well.  

\bigskip
\begin{center}
	{\large\bf SUPPLEMENTAL MATERIALS}
\end{center}
The R programming language was utilized for analyzing the data described in this paper, running simulations, implementing GAWS, and producing and comparing results, including for the pca+hdr and stray methods. All simulations, data and code have been made publicly available in a zipped file that can be found at the link below, with an accompanying readme.txt describing how to run all scripts. 
\newline
$https://github.com/colesodja/GAWS_LOCAL/blob/master/GAWSProject.zip$ \newline

$\textbf{Data}$: The saved time series simulations, pre-processed hourly pedestrian counts and transformed anonymized cloud traffic data are available as dataframes under a folder called data. 
\newline 

$\textbf{Simulations}$: If there is interest in generating additional sample paths then all the code to do so is located under a folder called sim. 
\newline 

$\textbf{Scripts}$: R code to re-produce results running the algorithms on saved datasets exists under the folder named scripts. 
\newline

$\textbf{Source Code}$: R functions for all GAMLSS models, penalized likelihood Akaike weight calculations, and the local GAWS algorithm is stored under the folder called src. Note that there are dependencies on other packages that must be installed. A script  $\textbf{1.install.packages.r}$ is included to check and if missing install all necessary packages. Here is a list of the packaged used: \newline 
$\textbf{gamlss}$ \newline
$\textbf{forecast}$ \newline
$\textbf{changepoint}$ \newline
$\textbf{splines}$ \newline
$\textbf{anomalous}$ \newline
$\textbf{oddstream}$ \newline
$\textbf{stray}$ \newline
$\textbf{sqldf}$ \newline

\end{document}